\documentclass[12pt]{iopart}
\usepackage{iopams}
\usepackage{graphicx}
\usepackage{epstopdf}
\usepackage{cite}
\usepackage{xcolor}
\usepackage{xspace} 
\usepackage{soul}

\newcommand{\HSEfamily}{{HSE($\alpha$, $\omega$)~}}

\newcommand{\HSE}[1][$\alpha$]{{HSE(#1)\xspace}}
\newcommand{\PBEh}[1][$\alpha$]{{PBEh(#1)\xspace}}

\newcommand{\EA}{EA\xspace}

\begin{document}

\title[Hybrid DFT for Fermi-level pinned systems]{Interface dipoles of organic molecules on Ag(111) in hybrid density-functional theory}

\author{Oliver T. Hofmann\footnote{Author to whom any correspondence should be addressed.}, Viktor Atalla, Nikolaj Moll\footnote{Permanent address: IBM Research -- Zurich, 8803 R\"uschlikon, Switzerland}, Patrick Rinke, and Matthias Scheffler}
        
\ead{hofmann@fhi-berlin.mpg.de, atalla@fhi-berlin.mpg.de, nim@zurich.ibm.com, rinke@fhi-berlin.mpg.de, scheffler@fhi-berlin.mpg.de}

\address{Fritz-Haber-Institut der Max-Planck-Gesellschaft, Faradayweg~4-6, D-14195 Berlin, Germany}

\date{\today}

\setcounter{footnote}{0}

\begin{abstract} 
We investigate the molecular acceptors 3,4,9,10-perylene-tetracarboxylic acid dianhydride (PTCDA), 2,3,5,6-tetrafluoro-7,7,8,8-tetracyanoquinodimethane (F4TCNQ), and 4,5,9,10-pyrenetetraone (PYTON) on Ag(111) using density-functional theory. For two groups of the \HSEfamily family of exchange-correlation functionals ($\omega= 0$ and $\omega = 0.2$\AA) we study the isolated components as well as the combined systems as a function of the amount of exact-exchange ($\alpha$). We find that hybrid functionals favour electron transfer to the adsorbate. Comparing to experimental work-function data, we report for $\alpha \approx 0.25$ a notable but small improvement over (semi)local functionals for the interface dipole. Although Kohn-Sham eigenvalues are only approximate representations of ionization energies, incidentally, at this value also the density of states agrees well with the photoelectron spectra. However, increasing $\alpha$ to values for which the energy of the lowest unoccupied molecular orbital matches the experimental electron affinity in the gas phase worsens both the interface dipole and the density of states. Our results imply that semi-local DFT calculations may often be adequate for conjugated organic molecules on metal surfaces and that the much more computationally demanding hybrid functionals yield only small improvements.
\end{abstract}


\maketitle
\tableofcontents

\title[]{} 

\section{Introduction}

A reliable theoretical description of the interaction between inorganic substrates and conjugated organic molecules is of fundamental importance for a variety of fields, including organic photovoltaics and (opto)electronics. For instance, the adsorption of strong electron acceptors creates an interface dipole modifying the work function of the inorganic substrate. Thus, a layer of acceptors can be inserted at a metal/organic interface to tune the charge-injection barriers into the organic material\cite{Ishii1999, Crispin2002, Koch2007}.  Here we focus on these so-called charge injection layers and investigate by means of density-functional theory (DFT) how molecular adsorbates affect the interface dipole formation.

In the framework of effective single-particle theories such as Kohn-Sham (KS) DFT, electron transfer from the substrate to the organic acceptor occurs through (partially) filling the lowest unoccupied molecular orbital (LUMO), which becomes pinned at the substrate's Fermi energy. The dipole that builds up as a result of the charge transfer increases the work function. However, most common exchange-correlation functionals, in particular (semi-)local functionals, suffer from noticeable electron self-interaction errors and the absence of the derivative discontinuity in the exchange-correlation potential. Subsequently, they place unoccupied orbitals too low and occupied orbitals too high in energy, which could result in spurious charge transfer.  This deficiency can be largely reduced by employing hybrid functionals, that include a fraction $\alpha$ of exact exchange. However, many hybrid functionals worsen the bulk properties of metals\cite{Paier2007}, which could adversely affect the description of the adsorption process.  We address this conundrum and investigate how the transition from (semi-)local to hybrid functionals affects metal/organic interfaces for two groups of the Heyd-Scuseria-Ernzerhof  (HSE)  \cite{Heyd2003} family.  HSE  builds on the Perdew-Burke-Ernzerhof (PBE) \cite{PBE0_1} generalized gradient functional and the corresponding hybrid functional PBEh  \cite{PBE0_1,PBE0_2,PBE0_3}.  In PBEh a fraction $\alpha$ of PBE exchange is replaced by exact (Hartree-Fock) exchange. A common choice is $\alpha=0.25$, more commonly known as PBE0 \cite{PBE0_1,PBE0_2,PBE0_3}. In HSE an additional range-separation parameter $\omega$ is introduced that limits exact exchange to the short range. $\omega= 0$ then corresponds to PBEh. Another common choice is the HSE06 parametrization that corresponds to $\alpha=0.25$ and $\omega = 0.2$\AA$^{-1}$ \cite{Heyd2003,Heyd2006}.

\begin{figure}
\includegraphics[width=\textwidth]{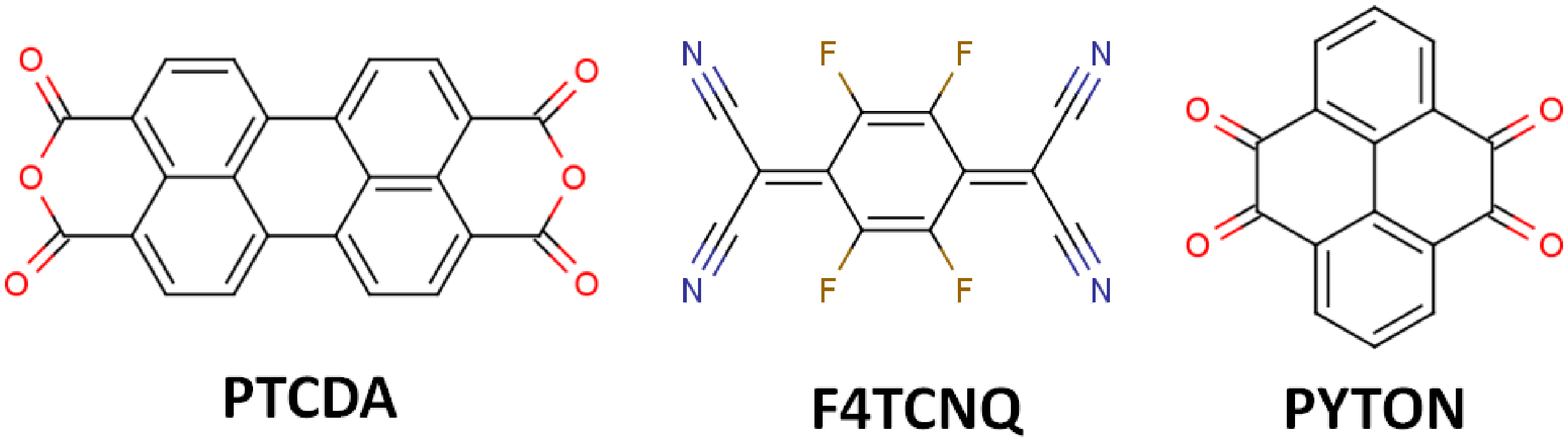}
\caption{\label{fStructures} From left to right: 3,4,9,10-perylene-tetracarboxylic acid dianhydride (PTCDA), 2,3,5,6-tetrafluoro-7,7,8,8-tetracyanoquinodimethane (F4TCNQ), and 4,5,9,10-pyrenetetraone (PYTON).}
\end{figure}

We have recently shown for pyridine on the non-polar ZnO($10\bar{1}0$) surface that the final work function depends strongly on the mixing parameter $\alpha$, while the interface dipole and therefore the adsorption induced work-function \textit{change} do not \cite{Hofmann/etal:2013}. However, there the interface dipole originates almost exclusively from the formation of a covalent bond.  We here extend our previous study to charge transfer systems. As test cases, we study PTCDA, F4TCNQ and PYTON (structures and full names are given in Fig.\ \ref{fStructures}) adsorbed on Ag(111), since for all three cases, experimental unit cells and work function (changes) as well as (semi-)local DFT results are available \cite{Tautz2007, Rangger2009, ThesisBen, PYTON_tobesubmitted}. In addition, the silver lattice constant differs by less than 0.2\% between PBE, HSE06, and PBE0 \cite{Marsman2008} and therefore we can keep the lattice constant fixed for all studies reported below.  

\section{Self interaction and hybrid density functionals}
\label{sSI-hyb} 

Organic/inorganic interfaces typically contain several 100 atoms per unit cell.  The first-principles method of choice for  systems of this size is KS DFT. (Semi)local DFT functionals are particularly popular, due to their computational efficiency. In a few notable cases  many-body perturbation theory \cite{Neaton2006, Freysoldt2009, Biller2011,Umari/etal_levelalignment:2013,Migani/etal:2013} or quantum Monte Carlo techniques have been used \cite{Zhigang/Kanai/Grossman:2009}, but due to their computational cost they still remain the exception. Furthermore, any perturbative treatment relies on the assumption that the electron density at the interface is well described by the zeroth order calculation, which is usually KS DFT. Thus any erroneous electron transfer that might occur in KS DFT is difficult to rectify in such a perturbative treatment \cite{atal+12tobe}.

\begin{figure}
\includegraphics[width=\textwidth]{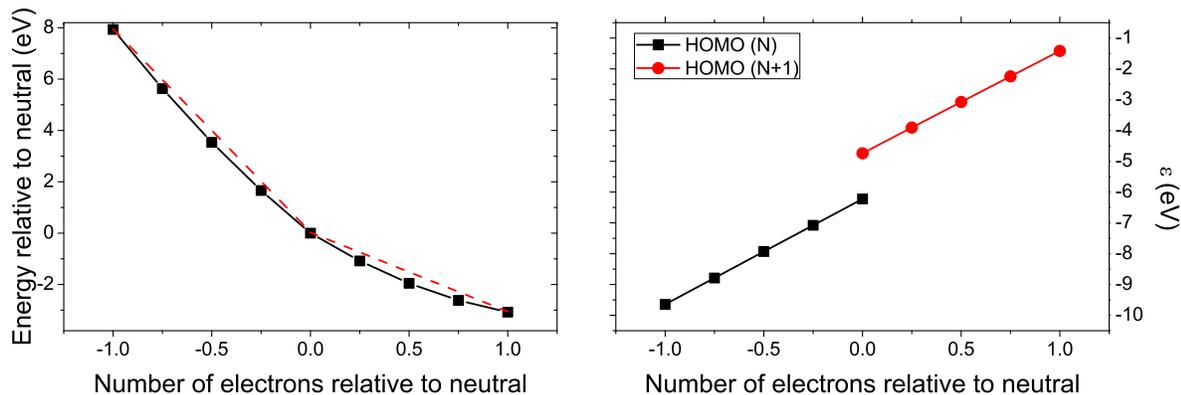}
\caption{\label{f_SI_PTCDA} Left: PBE total energy (black) for PTCDA in the gas phase with different charge states. $N=0$ corresponds to the neutral molecule. For comparison, the red dashed lines illustrate the ideal piecewise linear evolution.  Right: Orbital energies of PTCDA in the gas phase with different charge states. Black squares corresponds to the HOMO of the neutral molecule, red circles to the HOMO of the radical anion, i.e. the former LUMO of the neutral molecule. The dashed line corresponds to the ionization energy as determined by Dori et al.\cite{Dori2006} }
\end{figure}

The main deficiency of (semi-)local DFT functionals for interface calculations (apart from the missing image effects in the Kohn-Sham energies \cite{Neaton2006, Freysoldt2009, Thygesen2009})  is the self-interaction error \cite{Perdew1981, Mori-Sanchez2006}. This interaction of an electron with itself leads to a delocalization of electron density and a total energy that is no longer piecewise linear as a function of the electron number \cite{Perdew1982, Cohen2008, Cohen2012, Yang2012, Kraisler2013}. Concomitantly, the error is known as delocalization error \cite{Cohen2008, Cohen2012, Yang2012}, but is also referred to as many-electron self-interaction error \cite{Perdew1981, Mori-Sanchez2006} or non-Koopmans compliant error \cite{Dabo/etal:2010}.  We illustrate this behaviour for PTCDA and the PBE functional in Fig.~\ref{f_SI_PTCDA}. 

\textit{A priori} the deviation from the straight-line behaviour says little about the ionization energies of a molecule in the gas phase. The ionization potential (IP) and the electron affinity (\EA) are given by the total energy difference between the neutral system and the anion or the cation. This is known as the $\Delta$-self-consistent field ($\Delta$SCF) approach and performs well for atoms and small molecules also for local and semi-local functionals\cite{Gunnarsson/Lundqvist:1976, Vydrov2005, Rostgaard2010}. Thus, total energy differences are hardly affected by the self-interaction error. It should be noted, however, that vibronic effects are typically not included in $\Delta$SCF and will also not be included in our study.

In Kohn-Sham DFT, the orbital energies are given by the derivative of the total energy with respect to the particle number \cite{Janak1978}. The PBE total energy of PTCDA in Fig.~\ref{f_SI_PTCDA} exhibits a concave parabolic behaviour. As a result, the corresponding Kohn-Sham eigenstate energy changes linearly with occupation. For the neutral system ($N=0$ in  Fig.~\ref{f_SI_PTCDA}), the PBE KS eigenvalue then considerably underestimates the experimental value. This underestimation can have profound consequences when we bring two different subsystems into contact. In the limit of negligible chemical interaction this can give rise to an overestimation of charge-transfer \cite{Zhao2005} and underestimation of charge-transfer excitation energies \cite{Tozer2003}. In the worst case, spurious charge transfer results even at infinite separation \cite{Ruzsinszky2006}, if the highest occupied KS state of one subsystem lies above the lowest unoccupied KS state of the other \cite{Ruzsinszky2006,Sini2011, atal+12tobe}. 

Hybrid functionals are a popular approach to mitigate the self-interaction error, as we will show in more detail in Section~\ref{sGPM}.  In this work, we study the impact of exact exchange for hybrid functionals of the PBE hybrid (PBEh) form: 
\begin{equation}
E_{\rm xc}^{\rm PBEh}=\alpha E_{\rm x}^{\rm exact}+(1-\alpha) E_{\rm x}^{\rm PBE}+ E_{\rm c}^{\rm PBE} \quad .
\end{equation}	
$E_{\rm xc}$ is the exchange-correlation energy, $E_x^{\rm exact}$ the Hartree-Fock (i.e., exact) exchange energy, $\alpha$ the mixing parameter, $E_{\rm x}^{\rm PBE}$ the PBE exchange and $E_{\rm c}^{\rm PBE}$ the PBE correlation energy. Functionals in the Heyd-Scuseria-Ernzerhof  (HSE) family \cite{Heyd2003} additionally depend on a range-separation parameter $\omega$ that spatially splits the exchange energy into a long-range (LR) and short-range (SR) part
\begin{equation}
E_{\rm x}^{\rm HSE}=E_{\rm x}^{\rm PBEh, SR}(\omega)+E_{\rm x}^{\rm PBE, LR} (\omega) \quad.
\end{equation}
 At short range the exchange energy is given by PBEh and at long range by PBE, while the correlation energy is always given by PBE. The HSE family of functionals facilitates a smooth transition from the Perdew-Burke-Ernzerhof (PBE) \cite{PBE0_1} generalized gradient approximation ($\alpha$=0, $\omega$=arbitrary) to the PBE0  \cite{PBE0_1,PBE0_2,PBE0_3}  hybrid functional ($\alpha$=0.25, $\omega$=0) via the popular HSE06 \cite{Heyd2006} functional ($\alpha$=0.25, $\omega$=0.2~{\AA}$^{-1}$).  Limiting the range of exact exchange significantly reduces the computational time. However, for any non-zero value of $\omega$ the potential of any HSE functional will inherit the incorrect exponential asymptotic decay of PBE. In this work, we will focus on the impact of $\alpha$, but to clarify the impact of range-separation, we use two different groups of the HSE functional family, $\omega=0$ and $\omega=0.2$\AA$^{-1}$. While it was shown for several solids, including Ag, that HSE06 yields better structural properties and atomization energies than PBE (although the improvement over PBE0 is small)\cite{Marsman2008, Stroppa2008}, a survey of the work function of six different transition metals showed no systematic improvement\cite{Stroppa2008}. If this is also the case for systems with strong charge-transfer character is not {\it a priori} clear and will be the topic of this article. 

\begin{figure}
\includegraphics[width=\textwidth]{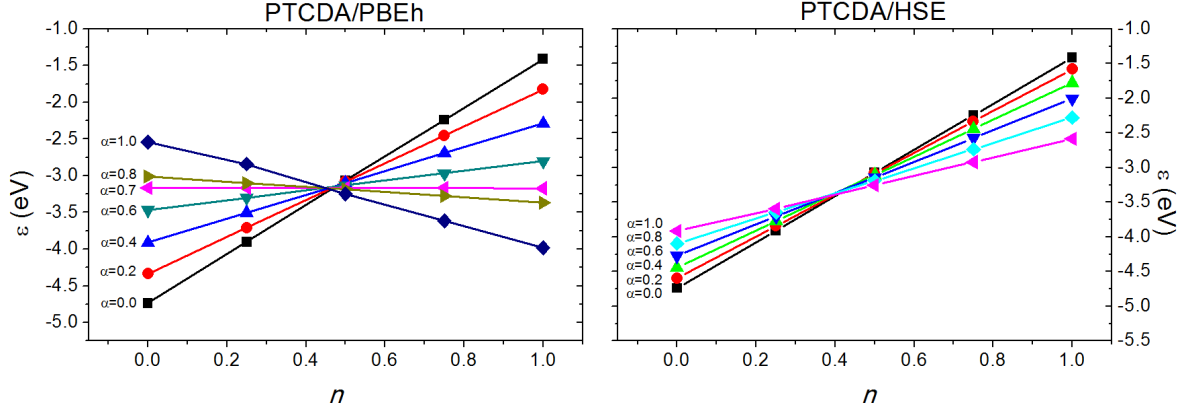}
\caption{\label{fFillings} LUMO orbital energy for PTCDA in the gas phase as function of its occupation, $n$, for different values of $\alpha$.}
\end{figure}

Applying the concept of the straight-line behaviour introduced earlier in this Section, we show in Fig.~\ref{fFillings} the variation of the  LUMO energy  ($\epsilon$) of PTCDA in the gas phase with respect to its orbital occupation. A many-electron self-interaction free description is reached when the orbital energy does not depend on its occupation, i.e. when it forms a straight, horizontal line. For \HSE, this criterion is never reached. Figure \ref{fFillings} shows a strong positive slope of $\epsilon$ for all $\alpha$, corresponding to a convex curvature of the energy \textit{vs.} occupation curve. For \PBEh, on the other hand, an almost perfect horizontal line is observed for $\alpha$ = 0.7. Smaller values of $\alpha$ show a positive, larger values a negative slope. The fact that all lines cross close to n=0.5 shows that the Slater-Janak transition state relation is fulfilled \cite{Slater1972, Janak1978}. The results also show that the Slater-Janak transition state is practically independent of the choice of the functional. 

For PTCDA (and later also for F4TCNQ and PYTON) we therefore conclude that {\PBEh} can be made self-interaction free and that this requires a large value of $\alpha$ = 0.8. In \HSE, on the other hand, the self-interaction error can never be fully removed. We attribute this to the fact that in \PBEh, exact exchange is not range limited \cite{PBE0_1}, which implies that the potential is closer to the exact $1/r$ behaviour, where $r$ is the distance from the molecule \cite{PBE0_1,PBE0_2,PBE0_3}. In \HSE, on the other hand, exact exchange is short ranged \cite{Heyd2003,Heyd2006} and the potential asymptotically follows the incorrect exponential decay of the PBE functional \cite{Vydrov2006}.  

We will, however, not dismiss the {\HSE} family at this stage, because we have not considered the metallic substrate, yet. For our surface/adsorbate system we therefore pose two questions. First, which $\alpha$ is \emph{best} for the combined system? And second, how do different hybrid functionals perform for the same $\alpha$? In the present study, we attempt to answer these questions by systematically investigating the effect of $\alpha$ for three different molecules adsorbed on Ag(111).

In contrast to earlier hybrid functional studies for physisorbed \cite{Sensato2002, Rimola2010, Biller2011} or covalently attached molecules \cite{Hofmann/etal:2013,Persson2000, Bludsky2005, Stroppa2008, Flores2009, Ren2009, Sala2010}, we here consider systems with strong charge-transfer character. We do this employing periodic boundary conditions. For cluster geometries, more hybrid functional studies exist (e.g. \cite{Sini2011, Ignaczak1997, Ignaczak1997a, Kua1999,  Repp2006, Santarossa2008, Fabiano2009, Li2012}, but they might miss collective electrostatic effects that develop due to the periodicity of the interface \cite{Deutsch2007, Rissner2012}.

\section{Computational details}
\label{sMethods} 

\begin{figure}
\includegraphics[width=\textwidth]{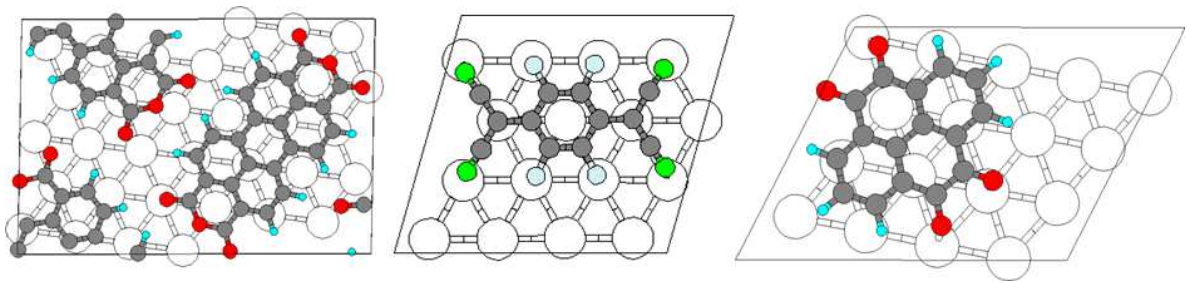}
\caption{\label{fUnitCells}  Left: Supercells for PTCDA, F4TCNQ, and PYTON on Ag(111). Only the top Ag-layer (white) is shown for clarity. Grey balls denote carbon atoms, cyan hydrogen atoms, blue fluorine atoms, red oxygen atoms, and green balls represent nitrogen atoms.}
\end{figure}

All calculations were performed with the Fritz Haber Institute {\it ab initio} molecular simulations (FHI-aims) code \cite{Blum2009}. Surfaces and interfaces were modelled by periodic slabs containing 5 layers of the metal. For surfaces, a region of  30~{\AA} vacuum was inserted between the metal slab and its periodic replica. The height of the unit cell was kept constant when adding the molecule, which causes a reduction of the vacuum layer by less then 5~\AA. Polarization  through the vacuum was prevented by means of the dipole correction \cite{Neugebauer1992}. Unless otherwise noted, all KS energies are reported with respect to the vacuum level. For non-periodic systems, this pertains to the potential energy of an electron at infinite distance. In periodic systems, the vacuum level is given by the electrostatic potential of an electron far above the surface. For practical purposes, here the energy 2\AA~beneath the place of the dipole correction is taken. 
 All computational parameters discussed below were converged for $\alpha = 0.8$ to a threshold of 10 meV for the work function. A ''Tier 2'' basis  was used for C, N, O, and F, which consist of the minimal basis plus one set of basis functions up to an angular momentum up to 3 ($d$-functions) and one set of basis functions up to an angular momentum of 5 ($g$-functions). The ''Tier 2'' basis for H consists of two sets of basis functions up to $p$ and $d$ functions, respectively. For Ag, only a single set of basis functions beyond the minimal basis, up to $f$-functions, were included. We verified that the removal of the $g$-function, which is part of the ''tight'' defaults, affects the total work function and the density of states by less than 10~meV, while significantly decreasing the computational demand and the memory requirements of the calculations. To obtain accurate surface dipoles, the cutoff potential of all basis functions was increased from 4~{\AA} to 6~{\AA}, which causes an increase of the work function of the pristine surface by approx. 50 meV. For integrations, a tightly converged Lebedev grid was used. A 35$\times$35$\times$1 off-$\Gamma$ k-point grid has been used for the primitive unit cell and scaled appropriately for the larger supercells. To account for van der Waals interactions, we employ the vdW$^{\rm surf}$  scheme \cite{Ruiz2012} for the metal-molecule interaction, which is appropriate for metal surfaces and has been shown to yield accurate adsorption distances and energies. Within this method, the van der Waals parameters for Ag are re-evaluated based on the dielectric function of the metal. Throughout this work, the values reported by Ruiz et al. \cite{Ruiz2012} were used. 
 The SCF-cycle was converged to a threshold of $10^{-4}$~eV for the total energy and $10^{-3}$~eV for the sum of eigenvalues. The unit cells were deduced from scanning tunnelling microscopy experiments \cite{Tautz2007, Rangger2009, PYTON_tobesubmitted}, and are shown in Fig.\ \ref{fUnitCells}. All geometries were optimized using the PBE+vdW$^{\rm surf}$ functional until the remaining forces were smaller than $10^{-3}$~eV/\AA. 

\section{Isolated molecules and the clean surface}
\label{sIS}

\subsection{Isolated gas phase molecules}
\label{sGPM}

As a first step, it is instructive to study the effects of exact exchange for the isolated subsystems. In principle, it would be best to compare DFT calculations to advanced computational methods, preferably coupled cluster singles double with perturbative triples (CCSD(T)), which is considered the "gold standard" of chemistry. Unfortunately, CCSD(T) calculations with a converged basis set are currently not available in the literature for charged, spin-polarized molecules of this size. For experimental reference values, on the other hand, we were  only able to find gas phase electron affinities for F4TCNQ.  Horke et al.\cite{Horke2011} measured the photoelectron spectrum of F4TCNQ anions at a photon energy resonant with the $D_2 \leftarrow D_0$ transition. By comparing to the spectrum of the unfluorinated derivative tetracyanoquinodimethane (TCNQ), they estimated an electron detachment energy of 3.2 eV. However, this analysis is aggravated by the fact that the $D_2$ state is below the vacuum level for F4TCNQ, but above the vacuum level for TCNQ. Neither extrapolation to T=0 K nor removal of zero-point vibration energies has been attempted.  It should be mentioned that the experimental ionization threshold may be affected by electron-vibrational coupling or the vibrational fine-structure. These effects are not accounted for in our calculations. Moreover, it is not always clear whether the ionization process in the experiment occurs vertically, i.e. at fixed geometry, or adiabatically. For these reasons, a perfect agreement between DFT calculations and experiment must not be expected. For PTCDA and PYTON, electron affinities in the  gas phase have not yet been measured to the best of our knowledge.

\begin{figure}
\includegraphics[width=\textwidth]{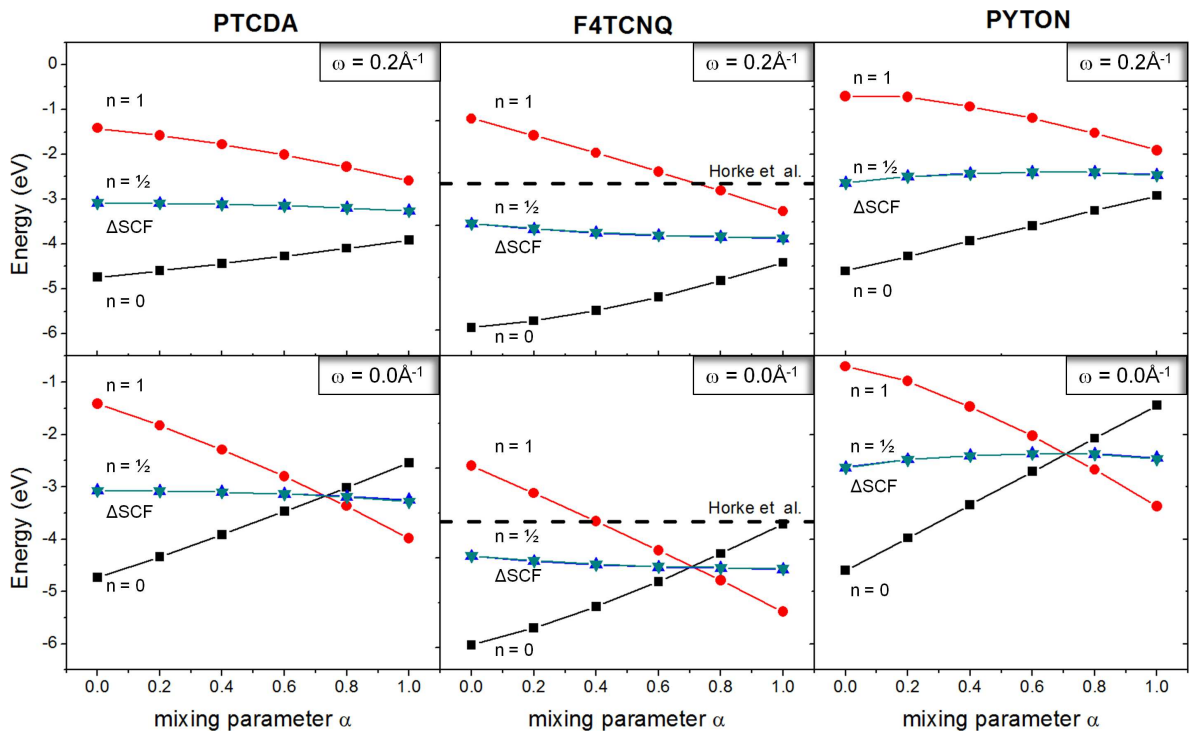}
\caption{\label{fSingleMoleculeEA} LUMO (black squares), SOMO (red circles), half-occupied orbital (green downward triangles) and and $\Delta$SCF-EA (blue upward trianlges) as function of alpha for PTCDA (left), F4TCNQ (middle) and PYTON (right). Top row: \HSE, bottom row: \PBEh. Experimental values (dashed lines) were taken from the literature \cite{Pshenichnyuk2011, Horke2011}. Note that $\Delta$SCF and n=$\frac{1}{2}$ are virtually on top of each other.}
\end{figure}

We first consider the electron affinity in  $\Delta$SCF, which is given by the ground-state energies $E_{\rm tot}$ of the charged ($N+1$ electrons) and uncharged systems ($N$ electrons):
\begin{equation}
{EA} = E_\mathrm{tot}(N) - E_\mathrm{tot}(N+1).
\end{equation} 
We only consider vertical electron affinities (i.e. the atomic positions were kept fixed in their equilibrium geometry of the neutral molecule) and compare the $\Delta$SCF-EA with the unoccupied LUMO ($n$=0) eigenvalues of the neutral molecule, the half ($n=\frac{1}{2}$), the and completely filled orbital ($n$=1) as a function of $\alpha$ for both hybrid functionals. The completely filled orbital corresponds to the HOMO of the $N$+1 electron system. In chemistry textbooks, it is commonly referred to as singly occupied molecular orbital (SOMO)\cite{Riedl}. We will adopt this notion here in order to avoid confusion between the HOMO of the $N$ and the $N$+1 electron system and to clearly distinguish the spin-polarized, partially filled orbital in the gas phase from the partially filled, spin-unpolarized orbital after adsorption (see below). 

Figure \ref{fSingleMoleculeEA} shows that the electron affinity in $\Delta$SCF lies almost exactly half way between the $n=0$ and $n=1$ and agrees with the $n=\frac{1}{2}$ energy. Thus the Slater-Janak transition state relation \cite{Slater1972, Janak1978} is fulfilled, as expected. We find that upon increasing $\alpha$, the energy of the unoccupied LUMO strongly increases with respect to the vacuum level, while the energy of the fully occupied orbital decreases. For the fully occupied or fully empty orbitals, the energies are significantly different when the same $\alpha$ but different $\omega$ is used. However, \PBEh\  and \HSE\  give virtually identical results for the half-filled orbital and $\Delta$SCF, in agreement with Ref. \cite{Krukau2006}, which reflects that the self-interaction error hardly affects these two approaches.

For atoms and small molecules, $\Delta$SCF typically performs well for ionization potentials and electron affinities already at the PBE level \cite{Gunnarsson/Lundqvist:1976, Vydrov2005, Rostgaard2010}. However, for F4TCNQ, we find that $\Delta$SCF gives electron affinities that are more than 1~eV lower than in experiment. Also $G_0W_0$ calculations \cite{Hedin1965}, that have become the method of choice for electron addition and electron removal energies in solids \cite{Rinke2008,Aulbur/Jonsson/Wilkins:2000}, do not provide agreement with experiment, as shown in \ref{sec:appA}. For the closely related molecule tetracyanoquinodimethane (TCNQ), CCSD(T) calculations with a small (aug-cc-pVDZ) basis set were reported\cite{Milian2004}. Even they cannot reproduce the experimental results, although significant improvement over $\Delta$SCF values was reported \cite{Milian2004}. The origin of this discrepancy is not yet understood. 

Returning to our definition of the many-electron self-interaction error in Section~\ref{sSI-hyb}, we observe that in our $\alpha$ dependent plots this condition is fulfilled when the energies for $n$=0, $n$=1, and $\Delta$SCF-EA ($n\sim\frac{1}{2}$) cross in the same point. For our three molecules, we find that for \HSE\ no such crossing point exists. Conversely, for the \PBEh\  functional the lines cross at $\alpha \approx 0.7$. In other words, for \PBEh\  a self-interaction free description of the orbital can be obtained, but not for \HSE. This is in agreement with our observations in Section~\ref{sSI-hyb}.

\begin{figure}
\includegraphics[width=\textwidth]{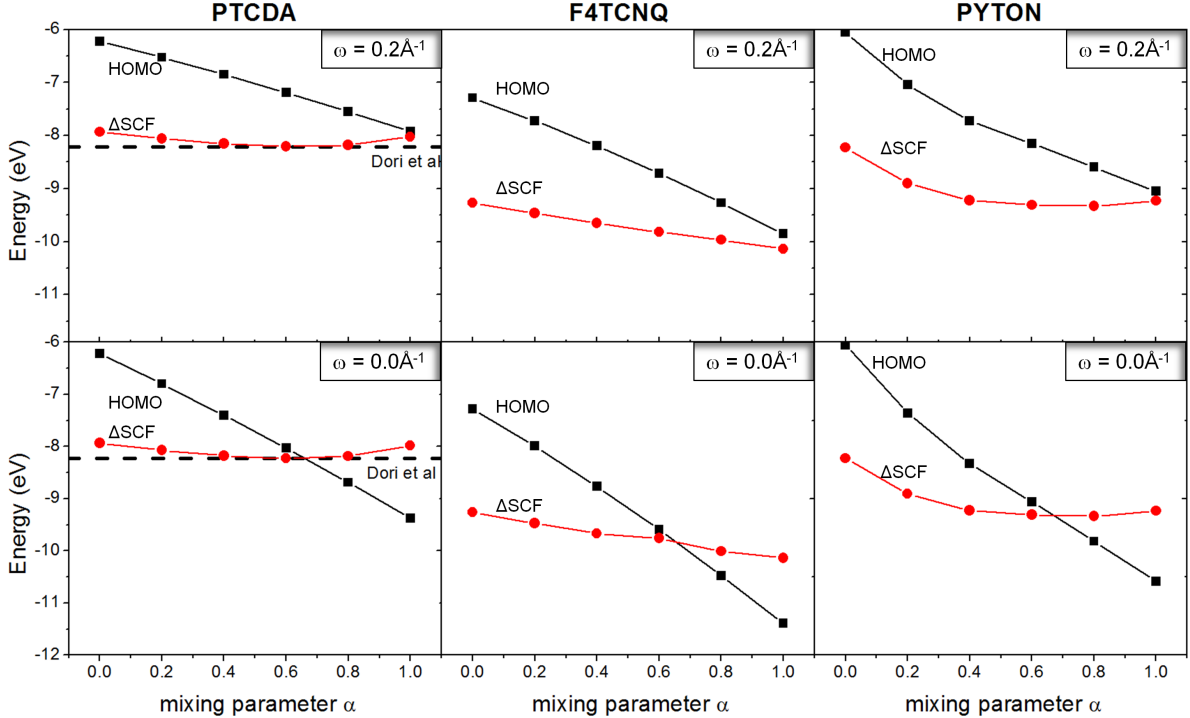}
\caption{\label{fSingleMoleculeIP} HOMO($N$) energy (black) and $\Delta$SCF, calculated as energy difference between the singly ionized and the neutral molecule gas phase. $\Delta$SCF agrees with the the KS-level at half occupancy ($\epsilon_N (\frac{1}{2}$). The experimental value (dashed lines) was taken from the literature. \cite{Dori2006}. }
\end{figure}

Figure \ref{fSingleMoleculeIP} shows a plot equivalent to  Fig.\ \ref{fSingleMoleculeEA} for the ionization potential (IP) of the molecules. In analogy to EA, the $\Delta$-SCF-IP was calculated as:
\begin{equation}
\rm{IP} = E_{\rm tot}(N-1) - E_{\rm tot}(N).
\end{equation} 

Experimental data for gas phase ionization is only available for PTCDA, for which Dori et al. determined an IP of 8.2~eV using photoelectron spectroscopy  \cite{Dori2006}. As above, no extrapolation to zero temperature or removal of zero-point energy was attempted. Nonetheless, here we find $\Delta$SCF and experiment to agree within 0.25 eV for any $\alpha$ and both $\omega$. Comparing KS eigenvalues with $\Delta$SCF, we encounter a similar situation as for the EA. For \HSE, the HOMO and the negative of the IP almost never agree, while for \PBEh, agreement in found close to $\alpha \approx 0.6$ throughout. This value of $\alpha$ is somewhat smaller than for the electron attachment energy, i.e.  HOMO and LUMO are thus affected differently by self-interaction.  An identical picture is obtained when comparing to perturbative $G_0W_0$ energies, see \ref{sec:appA}. 
For the organic molecules in the gas phase, we can now conclude that (a) the KS eigenvalues of the \PBEh\  group are better than those of the \HSE\  group and (b) generally large values of $\alpha$ are needed.

\subsection{The pristine metal surface}
Although it is in principle possible to perform $\Delta$SCF calculations for an embedded metal cluster without periodic boundary conditions, it is often challenging to ascertain whether a discrepancy between theory and experiment originates from deficiencies of the DFT functional or from finite size effects. For periodic systems, $\Delta$SCF calculations are aggravated by the fact that the addition of extra charge in the unit cell requires the introduction of a compensating charge to prevent the divergence of the electrostatic energy. The energy contribution from the compensating background then needs to be removed carefully.  For periodic surfaces, any homogeneous compensating background will introduce a dipole and thus an additional divergence with respect to an increase in the vacuum separation. Again, correction schemes exist \cite{Scheffler-87,komsa_finite-size_2013,Richter,Moll:2013}, but the physical energy contributions need to be carefully disentangled from those of the artificial, compensating background. We therefore refrain from $\Delta$SCF calculations for periodic surface supercells and determine $\alpha$ differently.

For semiconducting substrates, it has been suggested to make $\alpha$ dependent on the static dielectric constant\cite{Shimazaki2009,KollerArchive}, but that is not well defined for metal surfaces. Some of us have recently established a correlation between defect formation energies and the valence band width as a measure of the cohesive energy\cite{Ramprasad2012}. This can be used to determine $\alpha$ for semiconductors or insulators, provided accurate reference data is available for the valence bandwidth \cite{Ramprasad2012}. For metals, no such relation has been established, yet. Instead, we decided to focus on the work function and the density of states from periodic boundary calculations.

\begin{figure}
\includegraphics[width=\textwidth]{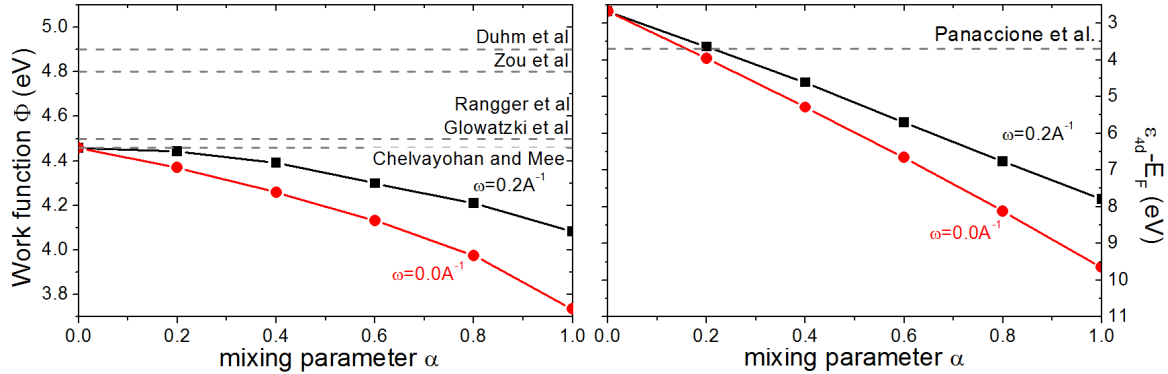}
\caption{\label{fAgAloneWF} Left: Ag(111) Work function $\Phi$ as a function of $\alpha$ for \HSE\  (boxes) and \PBEh\  (circles), compared to experimental results \cite{Rangger2009, Zou2006, Duhm2007, ThesisBen, PYTON_tobesubmitted}. Right:  Ag d-band onset as function of $\alpha$ for \HSE\  (boxes) and \PBEh\  (circles), compared to experimental results \cite{Panaccione2005}.}
\end{figure}

For Ag(111), the evolution of the Fermi-level with respect to the vacuum level (i.e., the work function $\Phi$) is shown in Fig.\ \ref{fAgAloneWF}a for a varying fraction of exact exchange. Both \HSE\  and \PBEh\  yield the highest work function for $\alpha = 0$, i.e. in the PBE limit. Increasing $\alpha$ results in a significant decrease of the work function of up to 0.4~eV (\HSE) and 0.8~eV (\PBEh). Stroppa and Kresse reported a similar decrease in work function for several different metals when going from PBE to HSE06, and ascribed it to a change in the surface dipole \cite{Stroppa2008} that results from a electron redistribution in the $d$ bands. 
Once again, ideally we would like to compare our results to higher-level theory. Unfortunately, coupled cluster calculations for extended surfaces are even less tractable than for molecules in the gas phase\cite{Booth2012}. The comparison with the experimental work function is aggravated by the fact that a wide range of values has been reported in the literature. These range from $\Phi = 4.46$~eV \cite{Chelvayohan1982}, and 4.5~eV (Ag(111) prior to F4TCNQ and PYTON deposition \cite{Rangger2009, PYTON_tobesubmitted}) to 4.8/4.9~eV (Ag(111) prior to PTCDA deposition \cite{Zou2006, Duhm2007}) as Fig.~\ref{fAgAloneWF}a shows. Despite this considerable spread, all experimental values are larger than the PBE work function. Such an underestimation of the work function is found for many metals \cite{Lang1971}. 

Although there is no requirement that the energetic position of the KS 4$d$-bands must agree with the corresponding photoemission peak, their incorrect description is often made responsible for artificial hybridizations. It is thus instructive for this work to also briefly study their dependence on $\alpha$. The onset of the Ag 4$d$ band is shown in the right panel of Fig.~\ref{fAgAloneWF}. The scale is one order of magnitude larger than for the work function. Between $\alpha = 0$ and 1, the $d$ band shifts by 5.1~eV (\HSE) or 7.0~eV (\PBEh). Unlike $\Phi$, the 4$d$ band is too high in energy when no exact exchange is present. This behaviour is well known for late transition metals \cite{MacDonald1982, Fuster1990}.  The calculations coincide with the experimental binding energy of 3.7~eV\cite{Panaccione2005} for relatively low fractions of exact exchange, $\alpha = 0.15$ to $\alpha = 0.20$ (\PBEh/\HSE), again in agreement with other studies \cite{Paier2006, Marsman2008}. We can thus conclude that (a) \HSE\  is more appropriate than \PBEh\  for the metal surface and (b), generally small values of $\alpha$ are needed.

\section{Fermi-level pinning for molecules on metallic surfaces}
\label{sflpmms}

\begin{figure}
\includegraphics[width=\textwidth]{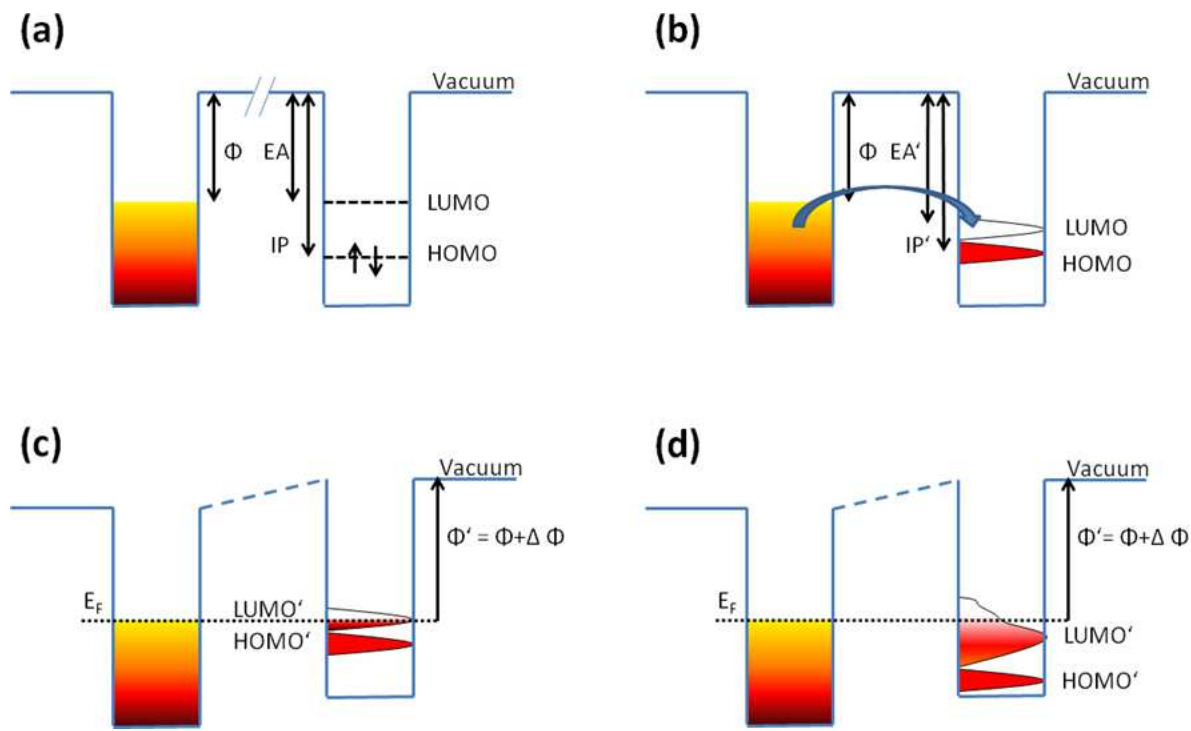}
\caption{\label{fFermilevelPinning}  (a) Hypothetical level alignment in the limit of non-interacting subsystems. (b) Upon contact, the discrete molecular levels broaden and electrons are transferred to the LUMO if it falls below the Fermi energy. (c) The electron transfer gives rise to an interface dipole. To avoid confusion, we denote the states of the adsorbate molecules HOMO' and LUMO', respectively. (d) Hybridization and molecular distortion may cause the density of states associated with the LUMO'  to deviate from a Lorentzian shape.}
\end{figure}

Before discussing the adsorbate systems in detail, it is instructive to briefly revisit the concept of Fermi level pinning. The basic idea behind Fermi-level pinning is illustrated in Fig.~\ref{fFermilevelPinning}. In the limit of large separation and no interaction between the subsystems, the metallic surface is described by its work function $\Phi$ (or equivalently, the Fermi-energy $E_{\rm F}$), whereas the organic adsorbate is characterised by its IP (HOMO) and EA (LUMO). Both systems share a common vacuum level (VL). Once the two are brought into contact, the molecular orbitals broaden (IP' and EA' in Fig.~\ref{fFermilevelPinning}). Also, the screening of charges near a metal surface allows the adsorbate to be more easily ionized, i.e. IP and EA are reduced \cite{Neaton2006, Freysoldt2009, Thygesen2009} (to IP' and EA' in Fig.~\ref{fFermilevelPinning}b). This effect is commonly referred to as band-gap renormalization or image effect. 

If the LUMO was not already below the Fermi level at infinite distance, it might be now. In that case the LUMO' becomes partially occupied as indicated in Fig.~\ref{fFermilevelPinning}(b). 
The electron transfer induces a shift of the adsorbate orbitals proportional to the dipole density\cite{Ishii1999}, $\mu/A$:
\begin{equation}
\Delta \Phi = \frac{1}{4 \pi \epsilon_0} \frac{\mu}{A}.
\end{equation}
Since in thermodynamic equilibrium the whole system must share a single Fermi energy, $\Delta\Phi$ will align the Fermi-energy with the LUMO'. In the end, the work function of the combined system, $\Phi '$, is given by the difference of the LUMO' to the vacuum level, see Fig.~\ref{fFermilevelPinning}(c). In real systems the situation is usually more complex, since hybridization and geometry distortions may lead to orbital broadenings which deviate significantly from the ideal Lorentzian shape and induce fractional occupations which pin the LUMO' level some tenth of eV away from the peak maximum, as sketched in Fig.~\ref{fFermilevelPinning}(d). Partially filled states at $E_{\rm F}$ associated with the LUMO' have been observed in the photoemission experiments for the three molecules discussed here \cite{Rangger2009, PYTON_tobesubmitted, Zou2006, Duhm2007}. 

\section{Level alignment in the non-interacting limit}
\label{slanl}

The previous section illustrates the importance of having a correct description of both the Fermi energy and the molecular frontier orbitals for Fermi-level pinned systems. At the same time, we have established that within \HSE~and \PBEh, the metallic surface and the organic adsorbate require very different values of $\alpha$. The obvious question is whether, and how, a compromise between those two different regimes can be found for a global value of $\alpha$. Before we examine the molecules adsorbed on the metal surface, however, it is insightful to examine the level alignment in the non-interacting-subsystem regime, in analogy to Fig.~\ref{fFermilevelPinning}(a). 

\begin{figure}
\includegraphics[width=\textwidth]{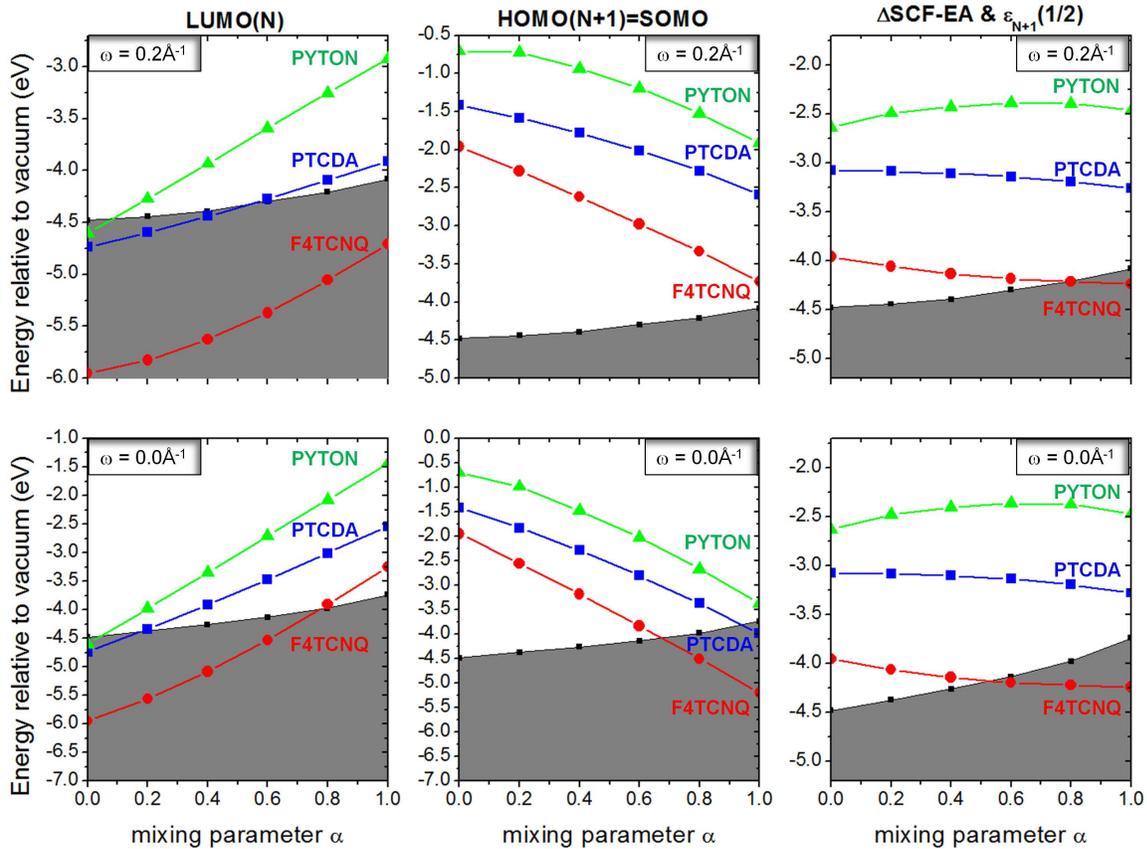}
\caption{\label{fAlignmentGasphase}  Level alignment in the non-interacting subsystem limit between the Fermi energy of Ag(111) and the molecular LUMO (left), SOMO (middle), and $\Delta$SCF-EA (right) in the gas phase. The top row shows \HSE\  and the bottom row \PBEh\  results. Blue squares denote PTCDA, red circles F4TCNQ, and green triangles PYTON.}
\end{figure}

For comparison we combine the data of the isolated molecules from Fig.~\ref{fSingleMoleculeEA} and the clean Ag surface from Fig.\ \ref{fAgAloneWF} into Fig.~\ref{fAlignmentGasphase}. For both functional groups, the empty LUMO crosses the Ag Fermi level for PTCDA and PYTON. For F4TCNQ, only PBEh exhibits a crossing of the LUMO and the Fermi energy. For PTCDA and PYTON, the crossing occurs at small $\alpha$. 
Similarily, for \HSE the filled SOMO is above the Fermi energy for all three molecules for any $\alpha$. In contrast \PBEh exhbits a crossing for PTCDA and F4TCNQ  at large $\alpha$. We also find that the $\Delta$SCF electron affinity lies significantly above the Fermi energy for both PTCDA and PYTON for any $\alpha$. Only F4TCNQ shows EAs which are partly below the Ag $E_{\rm F}$. These results convey the impression that PTCDA and PYTON on Ag(111) might not be Fermi-level pinned systems at all or that there is at least a transition from a charge-transfer to a non-charge-transfer regime, depending on the choice of $\alpha$. It should be kept in mind, however, that important factors are missing from this picture. On the one hand,  the electron density of the more polarizable component is partly displaced, as soon as two systems come into contact.\cite{Bagus2002}. This effect is known as Pauli pushback or pillow effect and significantly reduces the surface dipole. For Ag(111), reductions of 0.4-0.6~eV are commonly reported \cite{Ito1998, Huckstadt2006}. On the other hand, PYTON and F4TCNQ undergo noticeable distortions upon adsorption, leading to the formation of an intrinsic molecular dipole, which can affect the relative position of the molecular and surface levels. Even in the absence of such distortions, the $\pi$-electrons and the positive nuclei can build up a considerable local dipole moment that shifts all molecular levels relative to the surface \cite{Duhm2008}. Finally, the screening of charge on the molecule by the metal electrons is of course completely absent in the calculations of the isolated molecule in the gas phase. Indeed, these effects combined are strong enough so that we find Fermi-level pinning for all three molecules on Ag(111), as discussed in the next section.

\section{Electron transfer to acceptor molecules on metallic surfaces}
\label{schmms}

\begin{figure}
\includegraphics[width=\textwidth]{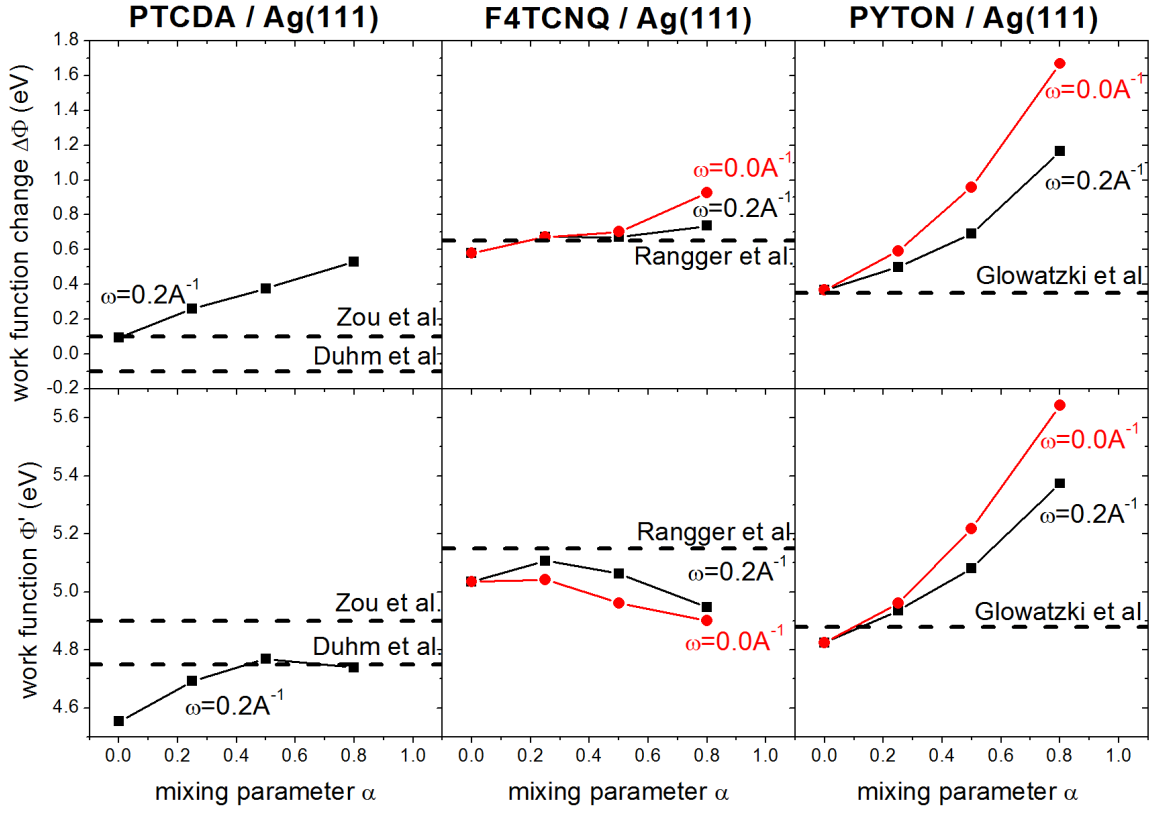}
\caption{\label{fWFChanges}  Top: Adsorption induced work-function modification, $\Delta \Phi$, for PTCDA, F4TCNQ, and PYTON (left to right) with \HSE\  (black boxes) and \PBEh\  (red circles). Bottom: Work function after adsorption of the molecules, $\Phi '$.  Note that bottom and top part of this Figure are directly related via Figure \ref{fAgAloneWF}a, i.e. $\Phi '=\Phi + \Delta \Phi$ }
\end{figure}

Despite the aforementioned deficiencies of standard (semi-)local functionals, remarkable agreement between the computed and the experimental work function has been reported for the adsorption of electron donors or acceptors on metal surfaces\cite{Crispin2002, Romaner2007, Romaner2009, Broker2008, Vazquez2007, Lindell2005, Lindell2008, Flores2009, Angela2010, Cakir2012}. Previously, this has been attributed to a fortuitous (partial) cancellation of errors \cite{Romaner2009}, in particular due to the fact that the HOMO-LUMO underestimation of (semi-)local functionals is of the same order of magnitude as the band-gap renormalization of the molecular states due to image effects that are absent from the PBE eigenvalues. Since hybrid functionals correct the former, but not the latter \cite{Biller2011}, the question arises whether the admixture of exact exchange improves or deteriorates the agreement between theory and experiment for the combined system. We focus first on the adsorption induced work function changes, $\Delta \Phi$, since this is a physical observable depending only on the electron density. Figure \ref{fWFChanges} shows $\Delta \Phi$ as function of $\alpha$.\footnote{We did not include \PBEh\  results for the largest system, PTCDA/Ag(111), because those calculations are significantly more expensive than \HSE calculations and could not be completed within a reasonable timescale.} 

For all three molecules, we observe a systematic, pronounced increase of the work-function change $\Delta \Phi$ with increasing $\alpha$. 
Using the two PTCDA/Ag(111) experiments as measure, we estimate the experimental reproducibility of work functions and work-function changes to be on the order of at least $\pm$0.2~eV. Comparing the experimental and the calculated work-function changes for PTCDA and F4TCNQ therefore yields good agreement for small and medium values of $\alpha$ (up to approx.\ 0.5). For PYTON, the work function change (i.e., the interface dipole $\Delta \Phi$) is more sensitive to $\alpha$ than for the other two molecules, and agreement with experiment is restricted to small values (approx.\ 0.25).  For the net work function of the system after adsorption (shown in the bottom part of the figure), the $\alpha$-induced reduction of the work function of the clean surface (cf. Fig.~\ref{fAgAloneWF}) and the increasing interface dipole partly compensate.  For PTCDA and F4TCNQ, the work function becomes almost independent of $\alpha$ as a result, varying by less than 0.25 eV. For PYTON, in contrast, the variation in $\Delta \Phi$ is much larger than in the other two cases, causing a strong net increase of $\Phi '$.  For the sake of brevity and clarity, we will show only the results for the \HSE\  group hereafter, since we find qualitatively similar results  for \PBEh\  and \HSE\  for both $\Phi '$ and $\Delta \Phi$.  

\begin{figure}
\includegraphics[width=\textwidth]{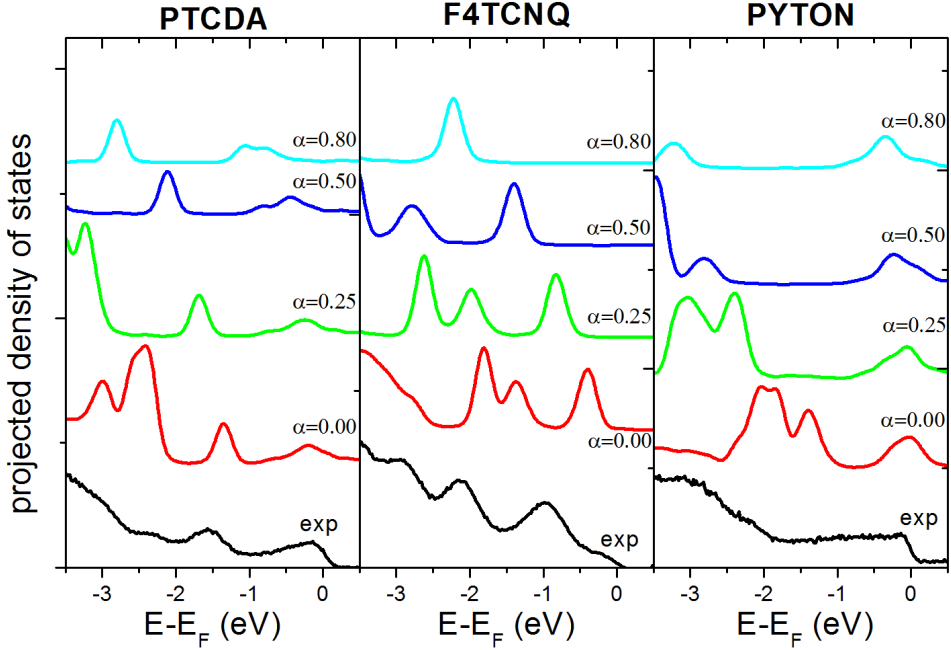}
\caption{\label{fDOS} Experimental photoelectron spectra (black) \cite{Rangger2009, Duhm2007, PYTON_tobesubmitted}  and projected density of states for PTCDA, F4TCNQ, and PYTON on Ag(111) (left to right) for different values of exact exchange $\alpha$. The spectra are vertically offset for clarity.}
\end{figure}

Since the geometry is kept fixed in our calculations, the change in $\Delta \Phi$ must be due to a change in hybridization or due to an increased electron transfer. To gain more insight, we turn to the density of states after adsorption which is shown in Fig.~\ref{fDOS} projected onto the respective molecule. Although a Kohn-Sham density of states is only approximately a physical observable,  K\"orzd\"orfer et al. pointed out that is is closely related to the experimental photoelectron spectrum\cite{Korzdorfer2009}, provided that electron-phonon coupling and disorder can be neglected\footnote{Note that the screening of the ionization energies by metal electrons affects occupied states of the same character and localization approximately equally\cite{Li2009} and thus hardly affects the energy difference of occupied $\pi$-states.}. If the aforementioned increase of the interface dipole is indicative of a larger electron transfer between metal and molecules, the molecular density of states should be shifted further below the Fermi level.  For all three molecules, we find the LUMO' to be below the Fermi energy and thus partly filled after adsorption. For $\alpha = 0$, the LUMO' resides too high compared to experiment. Increasing the fraction of exact exchange shifts the whole density of states to lower energies. At $\alpha \approx 0.25$, the LUMO' agrees well with the photoemission experiments. A further increase of $\alpha$ shifts the LUMO' to even lower energies. The LUMO' therefore behaves qualitatively like the SOMO, rather than the LUMO of the isolated molecule (cf. Fig.~\ref{fAlignmentGasphase}), except for the fact that it agrees with experiment at much smaller fractions of exact exchange. The observed downward shift causes an increased filling of the LUMO' (see below for more details), and thus an increased electron transfer. We attribute the reason for this behaviour to the fact that for all molecules, the LUMO at $\alpha = 0$ is occupied with more than one electron (see below for details), and exact exchange shifts orbitals down in energy if they are more than half filled and down otherwise (cf. Figure \ref{fFillings}).  We therefore speculate that the general trend of increased electron transfer only holds for system for which the LUMO' is more than 50\% filled at the PBE level, and that the reverse (i.e. reduced electron transfer) should be observed for systems with less than half filling at the PBE level. 

\begin{figure}
\includegraphics[width=\textwidth]{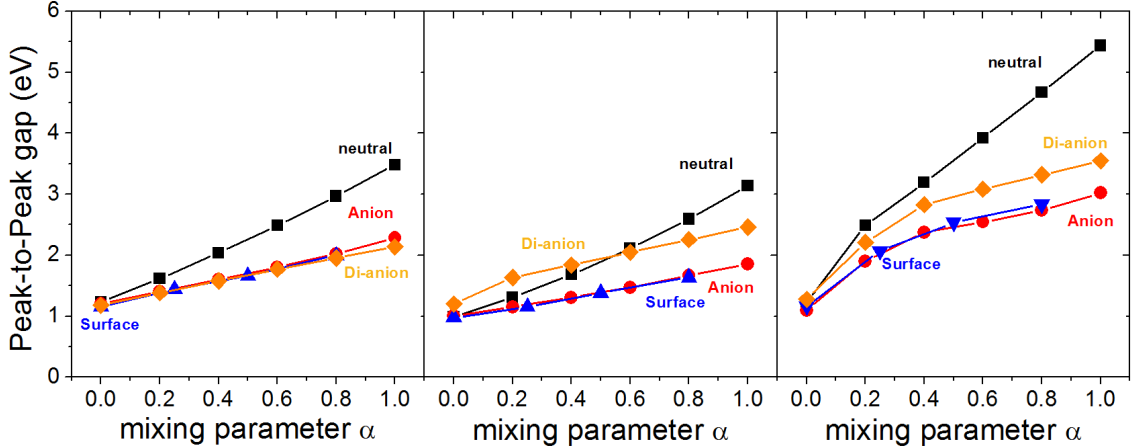}
\caption{\label{fGaps} Generalized KS gap between LUMO' and HOMO' on the surface (blue triangles), the gas phase molecule in its neutral state (black square), the singly charged anion (red circles) and the di-anion with a doubly occupied LUMO' (orange diamonds).}
\end{figure}

The $\alpha$-induced shift is not the same for every orbital. In fact, the HOMO' shifts more than the LUMO', causing an increase in the HOMO'-LUMO' splitting. Interestingly, however, Fig.~\ref{fDOS} reveals that both orbitals are simultaneously in excellent agreement with experiment at $\alpha \approx 0.25$. At this point, we briefly digress to emphasize another interesting finding. Figure \ref{fGaps} shows the evolution of the gas-phase HOMO-LUMO gap of the neutral molecule, the energy difference between the SOMO and the next lowest orbital in the same spin channel (former HOMO), the energy difference between the doubly occupied LUMO' and HOMO' of the molecular di-anion, and the HOMO'-LUMO' gap deduced from the adsorbed molecules on the surface ($E'_{\rm gap}$). To exclude geometry effects, the orbital energies were calculated using the geometry that the molecules assume on the surface after adsorption. The calculation for the di-anion was calculated spin restricted, i.e., for all molecules we report the singlet rather than the triplet state. For the surface calculations, we determine the gap as peak-to-peak rather than onset-to-onset, since we used an artificial broadening (see Section \ref{sMethods}). Furthermore, the experimental broadening  -- and thus the onset-to-onset gap -- is partly determined by electron-phonon coupling of the states and disorder, which is not included in our calculations \cite{Tautz2002, Kilian2008, Ciuchi2012, Patrick2012}.

On the surface, $E'_{\rm gap}$ increases much slower than the gap between the HOMO and the LUMO in the gas phase. $E'_{\rm gap}$ does also not follow the behaviour of the the gas phase di-anion despite the fact that at the surface the LUMO is almost completely filled with two electrons (see next paragraph). In contrast, $E'_{\rm gap}$  closely tracks the $\alpha$-dependence of the radical anion in the gas phase. This result is rather unexpected, not only because the occupation of the LUMO on the surface differs significantly from 1, but also because the gas phase anion is an open shell system and was treated spin polarised, while the adsorbate calculation was not spin-polarised. At present, it is not clear whether this result is coincidental or not, as correlation does not necessarily imply causation. Nonetheless, this ambiguity might have contributed to the controversy regarding the contentious issue of integer (a single electron to a fraction of molecules) versus fractional charge transfer (fractional charge to all molecules), that is particular well documented for F4TCNQ \cite{Soos2011}. 

\begin{figure}
\includegraphics[width=\textwidth]{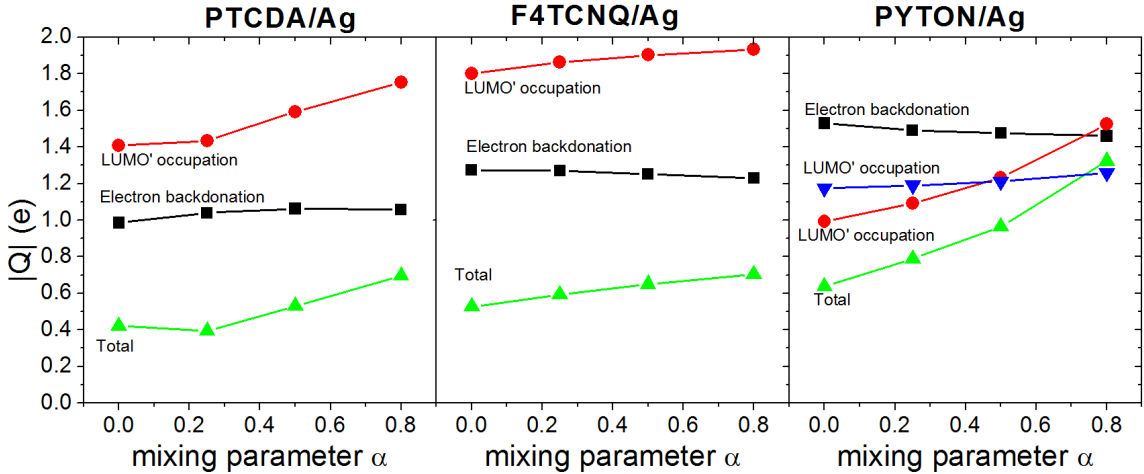}
\caption{\label{fMODOS} Absolute values of the total charge transfer (green triangles), LUMO occupation (red circles), and charge donation (black squares) for the three test molecules on Ag(111). For PYTON the LUMO is doubly degenerate in the gas phase and we therefore show both states (red circles and blue triangles).}
\end{figure}

To better understand the evolution of $\Phi '$ and $\Delta \Phi$ in terms of the properties of the isolated subsystems, it is important to review the charge-transfer mechanism in more detail. To this aim, we project the density of states onto the molecular orbitals of the hypothetical, free-standing monolayer (MODOS) \cite{Nelin1987}. By integrating the molecular density of states of each orbital from $-\infty$ to $E_{\rm F}$ we then obtain formal occupations for each orbital. This allows us to quantify donation and back donation separately. Here, we define donation as the charge accumulated by the LUMO, and back donation as the total charge of the molecule minus the donation. Although this definition differs somewhat from the more intuitive definition in which donation is given by the sum over the occupation numbers of all occupied orbitals and the back donation as the sum over all unoccupied levels, it has the advantage of being more easily identifiable with the experimental photoelectron spectra, for which the LUMO' occupation is evident. To be consistent with earlier work \cite{Rangger2009}, we employed a Mulliken-like projection. The results are therefore subject to the corresponding shortcomings, such as occupations potentially exceeding the 0 to 2 range and an increasing ambiguity with diffuse basis sets. However, it has been found earlier that the general trends obtained by this analysis compare well with trends obtained, e.g., by real-space integration of the electron density \cite{Hofmann2008}. Our results for the net charge transfer, as well as donation and back donation are shown in  Fig.~\ref{fMODOS} (note that the absolute values of all charges are plotted). 

For PTCDA, we find a considerable net charge-transfer between PTCDA and Ag(111), in agreement with earlier work \cite{Tautz2002, Hauschild2005, Duhm2008, Romaner2009, Schwalb2010}. Using our charge-partition scheme, the charge transfer amounts to approx.\ 0.4~$e$ for $\alpha = 0$, which increases gradually to approx.\ 0.7~$e$ for $\alpha = 0.8$. The origin of the increased charge transfer can be traced back to an increased occupation of the LUMO', which rises from 1.4~$e$ to 1.7~$e$. Charge donation, on the other hand, remains almost constant at approximately one electron. The increased charge transfer then gives rise to an increased interface dipole. 

F4TCNQ interacts with Ag(111) trough the hybridization of the low-lying CN-orbitals with the Ag $d$-bands \cite{Rangger2009} and the filling of the LUMO'. Already for $\alpha = 0$, the LUMO' is found almost completely below the Fermi energy, giving rise to an occupation of 1.8~$e$. Exact exchange further stabilises the now occupied orbital, shifting it to lower energies. However, since the LUMO' is already almost completely filled at the PBE-level, this shift only causes a minor increase of the occupation to 1.9~$e$. At the same time, charge donation remains unaffected at a level of 1.3~$e$. The total amount of charge transfer therefore remains essentially constant, increasing only very little as $\alpha$ increases. As a result, the net work function of F4TCNQ on Ag(111) follows the work-function evolution of the pristine Ag surface.

In the gas phase, PYTON has a degenerate LUMO. The adsorption induced distortion and the interaction with the threefold symmetric surface lift this degeneracy, albeit only by some tenth of an eV. The hybridization-induced broadening of these orbitals is larger than this split and their densities of states continue to overlap. As a result, both orbitals become occupied upon adsorption. Although both orbitals are $\pi$-orbitals, they are not equally localized and thus differently affected by self-interaction. This leads to a reordering of these orbitals on the surface when $\alpha$ increases. While the position of one orbital is essentially unaffected, resulting in a constant occupation of approx. 1.2 e, the other orbital shifts strongly with $\alpha$ to more negative energies, and thereby increases its occupation from 1.0 to 1.5 e. This molecule therefore accepts up to 2.7 electrons in total. Naturally, the large electron donation must be accompanied by a large electron back-donation. Indeed, we find a back donation of approx.~1.5 electrons, originating from the interaction of the carbonyl-states with Ag(111).  As for PTCDA and F4TCNQ, the donation is only weakly dependent on the choice of $\alpha$. Once again, the net result is a strongly increasing charge transfer, which agrees with the reported strong increase of the work function with $\alpha$. 

The overarching  question that remains, is why the three different molecules react so differently to changes in $\alpha$, i.e., why the interface dipole increases much more for PYTON than for PTCDA, and for PTCDA more than for F4TCNQ. We could establish no simple relation between the $\alpha$-dependent $\Phi'$ or $\Delta \Phi$ and the electron affinity, the LUMO, or the SOMO, neither for the gas phase equilibrium structure nor the structure including adsorption-induced geometry changes. We thus have to conclude that the generalized KS eigenvalues of the gas phase molecules alone are in general not reliable indicators for Fermi-level pinning and cannot be used to predict the size of work function changes. Nonetheless,  Fig.~\ref{fMODOS} clearly shows that the influence of $\alpha$ on charge back donation is negligible. This in turn implies that any change in the charge transfer, which manifests itself in the interface dipole, must be related to changes in the LUMO' position. We can therefore identify two system-dependent variables. First, the position of the LUMO' for $\alpha$=0. In contrast to the situation on Cu or Au, where a complete filling only occurs at low coverages \cite{Soos2011, Topham2011}, the LUMO' is almost completely filled for F4TCNQ even at full coverage on Ag(111). Further $\alpha$-induced downwards shifts can therefore not yield a significant change in charge transfer. PTCDA and PYTON, on the other hand, are only partially filled for $\alpha = 0.0$ and can still accept more charge. Moreover, the change in the transferred charge is  directly proportional to the molecular density of states at $E_{\rm F}$  and therefore depends on the area of the LUMO' peak that is shifted below $E_{\rm F}$. As Fig.~\ref{fDOS} shows, the density of states for any given $\alpha$ is always largest for PYTON and smallest for F4TCNQ. The exceptionally large DOS at $E_{\rm F}$ is a consequence of the near-degeneracy of the two LUMOs and explains the extraordinarily large work function (change) of PYTON compared to the other molecules. This observation is in line with the Induced Density of Interface States (IDIS) model \cite{Vazquez2004}, which emphasizes the importance of the adsorption-induced molecular density of states for systems of this kind.  A second factor is clearly the sensitivity of the orbital eigenvalues to $\alpha$, which depends on the orbital's localization and its filling. Figure \ref{fSingleMoleculeEA} shows that the PYTON orbitals are much more sensitive to exact exchange than those of PTCDA. This gives a larger downward shift of the newly occupied orbital. Both aspects together form a plausible, if not comprehensive explanation for the observed variation of $\Delta \Phi$ and the charge-transfer dependence of the three systems.

\section{Summary and conclusions}
\label{ssc}

All the above findings may confer the impression that the adsorption of Fermi-level pinned, conjugated organic molecules on metal surfaces is already well described  at the PBE+vdW level, and that improvements obtained from \HSE\ or \PBEh\ are hardly worth the significantly higher computational effort. Indeed, as a general \emph{rule of thumb} we expect this to be true. Nonetheless, it is clear that some systems react more sensitively to hybrid functional than others, for example when they exhibit a particularly high density of states at the Fermi-energy as the case of PYTON demonstrates. We thus expect molecular adsorbates that have degenerate frontier orbitals to require a beyond PBE+vdW treatment. It should also be kept in mind that we only considered molecules that bind to the metal surface through hybridization and donation/back-donation. It is not clear yet what will happen in the absence of hybridization, especially if charge-transfer to the organic layers causes spin-polarization. Last but not least, all our calculations place the same charge on all molecules in the layer,  because of the periodic boundary conditions. If, however, charge transfer becomes localised on one out of many molecules, corresponding lattice distortions (polarons) may become important as reported for both polymers and small organic molecules \cite{Braun2009}.

In summary, we have systematically studied the level alignment of three different conjugated organic molecules on Ag(111) as function of the exact exchange admixture $\alpha$ in the hybrid functionals \PBEh\  and \HSE. The orbitals of the isolated molecules in the gas phase require a large fraction of exact exchange ($\alpha \approx 0.7 {\rm \ to \ } 0.8$) to become self-interaction free within the \PBEh\  group, while for \HSE\  this condition is never fulfilled. In contrast, the work function of the clean Ag surface is always underestimated by all hybrid functionals, whereas an accurate position of the $d$-bands with respect to the Fermi energy is obtained for $\alpha \approx 0.25$. In general, we find that screened exchange (\HSE) works better than  \PBEh\  for Ag. 

For the combined inorganic/organic systems, we find that the interface dipole upon adsorption is best described at $\alpha \approx 0$, whereas the net work function after adsorption requires $\alpha \approx 0.25$. The photoelectron spectra, in particular the HOMO' and LUMO' positions, are best reproduced for $\alpha \approx 0.25$. For the three systems studied here, more exact exchange leads to a systematic increase in charge transfer, which we attribute to the fact that all molecules pin at the Fermi-level and exhibit a large LUMO' occupation already at the PBE level. This can be directly related to an increase of the LUMO' occupation, whereas charge donation remains almost unaffected. However, overall the difference between $\alpha = 0$ and $\alpha = 0.25$ is relatively small. Given the fact that the reproducibility of organic monolayers on metal surfaces is still low and that experimental results exhibit a certain spread, we conclude that PBE-based results for interface dipoles and densities of states are generally adequate for interfaces between coinage metals and conjugated organic molecules. As the example of PYTON shows, tuning $\alpha$ to match orbital eigenvalues and ionization potentials or electron affinities can be even counterproductive for global and short-ranged range-separated hybrid functionals.

\addcontentsline{toc}{section}{Acknowledgments}
\section*{Acknowledgments}

We would like to thank Steffen Duhm, Benjamin Br\"oker, Hendrik Glowatzki, and Norbert Koch for providing the experimental photoelectron spectra of PTCDA, F4TCNQ and PYTON, Victor Ruiz-Lopez for providing the PTCDA structure, and Sergey Levchenko, Volker Blum, Heiko Appel, Georg Heimel, and Egbert Zojer for fruitful discussions. Funding through SFB 951 (HIOS) and the FWF project J 3285-N20  is  gratefully acknowledged.   

\appendix

\section[\hspace{2cm}Quasiparticle energy calculations for gas phase molecules]{\label{sec:appA}Quasiparticle energy calculations for gas phase molecules}

\begin{figure}
\includegraphics[width=\textwidth]{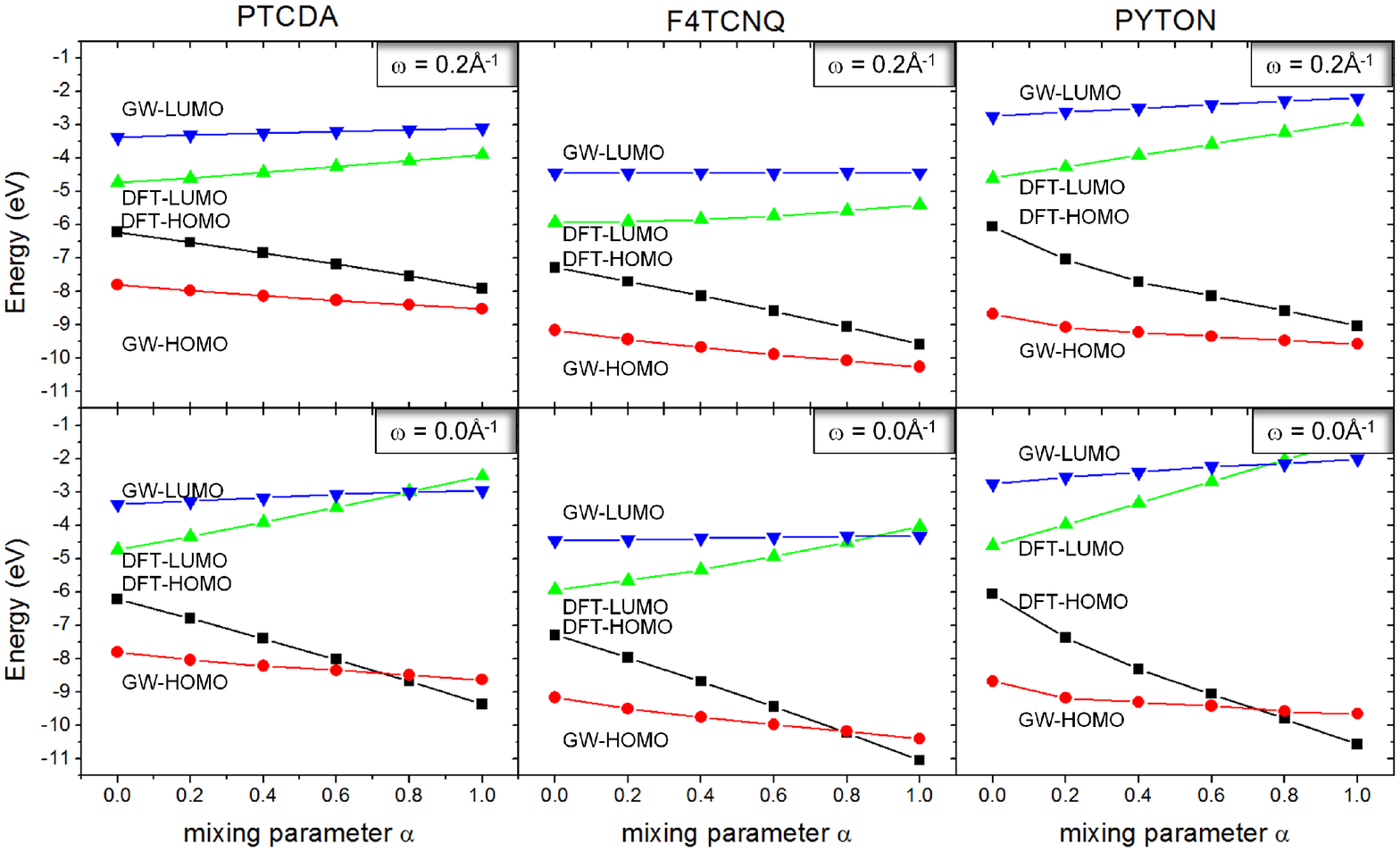}
\caption{\label{fSingleMoleculeGW} HOMO and LUMO of PTCDA, F4TCNQ and PYTON in base phase for the two hybrid functionals and  $G_0W_0$@hybrid functional as a function of $\alpha$. }
\end{figure}

In Figs.\ \ref{fSingleMoleculeEA} and \ref{fSingleMoleculeIP} we have compared orbital eigenvalues to energy differences of the same functional. However, it is beneficial to also compare to a more sophisticated method. As alluded to in the introduction, for solids Hedin's $GW$ approach \cite{Hedin1965} has become the method of choice for the calculation of quasiparticle band structures \cite{Aulbur/Jonsson/Wilkins:2000,Rinke2008}. $GW$ is also becoming increasingly more popular for molecules. Since fully self-consistent $GW$ is computationally still too expensive \cite{Caruso/etal:2012,Caruso/etal:2013_tech}, $GW$ is commonly applied perturbatively ($G_0W_0$). Due to $G_0W_0$'s starting-point dependence \cite{Rinke/etal:2005,Fuchs/etal:2007,Marom2012,Bruneval/Marques:2013} we apply $G_0W_0$ to all our hybrid functional calculations. We use the $G_0W_0$ implementation of FHI-aims \cite{Xinguo/implem_full_author_list} and Tier 3 basis sets. The results are shown in Fig.~\ref{fSingleMoleculeGW}.   

The LUMO is almost independent of $\alpha$ in $G_0W_0$, while the HOMO exhibits a starting-point dependence of the order of 1~eV for the three molecules. We again observe that the \HSE\  eigenvalues never cross the $G_0W_0$ lines, whereas the \PBEh\  HOMO crosses at $\alpha \approx 0.7$ and the LUMO at $\alpha \approx 0.8$. These values are consistent with our observations in Section~\ref{sGPM}. In principle we could use the intersection between $G_0W_0$ and \PBEh\  to design an internally consistent $G_0W_0$ scheme \cite{atal+12tobe}. The highest KS eigenvalue of a finite system is given exactly in exact DFT \cite{Almbladh1985,Levy/Perdew/Sahni:1984} and its self-energy correction vanishes. This would be the intersection of the DFT-KS and the $G_0W_0$ line. We here rely on the fact that this relation is also fulfilled for approximate DFT functionals and the $GW$ self-energy. The intersection then gives us an optimised $\alpha$ value. Since it also determines an internally consistent starting point for $G_0W_0$, the scheme can be viewed as a simplified self-consistent $GW$ procedure \cite{atal+12tobe}. This internally consistent scheme is of course also applicable to solids and interfaces. However, currently $G_0W_0$ is not yet implemented for periodic boundary conditions in FHI-aims.


\addcontentsline{toc}{section}{References}
\section*{References}

\bibliographystyle{iopart-num}
\bibliography{Reference}

\end{document}